\documentclass[doublecol,figures]{epl2} 

\usepackage{amssymb}

\newcommand{\dervi}[2]{{\rm d} #1/{\rm d} #2}

\title{Observation of quasi-periodic frequency sweeping in electron cyclotron emission of nonequilibrium mirror-confined plasma}
\shorttitle{Observation of quasi-periodic frequency sweeping in electron cyclotron emission}

\author{M. E. Viktorov\thanks{E-mail: \email{mikhail.viktorov@appl.sci-nnov.ru}} \and A. G. Shalashov \and D. A. Mansfeld \and S. V. Golubev}
\shortauthor{M. Viktorov \etal}
\institute{ Institute of Applied Physics of Russian Academy of Sciences -- 46 Ulyanov Str., 603950, Nizhny Novgorod, Russia}


\abstract{
Chirping frequency patterns have been observed in the electron cyclotron emission from strongly nonequilibrium plasma confined in a table-top mirror magnetic trap. Such patterns are typical for the formation of  nonlinear phase space structures in a proximity of the wave-particle resonances of a kinetically unstable plasma, also known as the ``holes and clumps'' mechanism. Our data provides the first experimental evidence for acting of this mechanism in the electron cyclotron frequency domain. 
}

\begin{document}

\maketitle

\section{Introduction}

Resonant interaction of electromagnetic waves and charged particles plays an important role in the dynamics of magnetoactive plasma confined in space and laboratory magnetic traps. One of the most intriguing manifestations of such interaction is emission of broadband  radiation  with regular variations of dynamical spectra, e.g. quasi-periodic bursts with a frequency sweeping, resulting from the development of kinetic plasma instabilities. Such events are common features of experimental plasmas. 
Kinetic instabilities are caused by the presence of positive gradients in the velocity distribution of resonant particles, whose formation is universal for both space and laboratory plasma. In space magnetic traps, 
the sources of free energy are formed due to different acceleration mechanisms of particles, such as betatron acceleration, plasma-wave turbulence and magnetic reconnection. Under laboratory conditions, the energetic particles with an anisotropic velocity distribution can be formed due to the features of plasma heating, when the energy of the external source is embedded in a specific region of the phase space, such as provided by resonant cyclotron heating or neutral beam injection in magnetic fusion experiment. 
{Spatial gradients can also result in instabilities of waves for which the diffusion in real space is coupled to the diffusion in velocity space due to conservation of invariants of particle motion in inhomogeneous systems energy; this type of instability is exploited in the so-called alpha channeling proposals \cite{Fisch1,Fisch2}.}
As a rule, plasma confined in laboratory traps consists of at least two components, one of which, more dense and colder, determines the dispersion properties of the unstable waves, and the second small group of energetic particles with an essentially anisotropic velocity distribution which is responsible for the growth of the  waves. 

One common description for the self-consistent evolution of particles and waves is the quasilinear theory, a perturbative approach that involves many overlapped wave-particle resonances as a basis for a diffusive particle transport in phase space\footnote{{Resonant interaction between monochromatic wave and particles in inhomogeneous magnetic field may also result in quasi-linear diffusion in momentum space, that happens because of the stochastic phase of particles during repeated passing trough the resonance region, see e.g. \cite{Tok}. Such concept of quasi-linear diffusion is widely used related to a high-frequency heating of fusion plasmas \cite{Gir1,Gir2}.}} \cite{ql1,ql2}. Within this approach, which is most developed for the whistler and Alfven wave instabilities in space \cite{trakh_book_2008}, the frequency sweeping may be explained as a result of relatively slow modification of the average distribution function of the resonant particles. Although the quasilinear theory is in principle able to describe fast events, such as switching from kinetic to {hydrodynamic} instability in the inner magnetosphere of the Earth {\cite{trakh1,trakh2,trakh3}}, it faces essential difficulties in many cases involving fast transients.  

An alternative universal physical mechanism is based on formation of nonlinear phase space structures in the proximity of the wave-particle resonances of a kinetically unstable bulk plasma {mode. In collisionless limit, such regimes were  proposed as a possible mechanism of generation of narrowband chorus emissions in the Earth's magnetosphere  \cite{trakh4,trakh5,Demekhov}. Evolution of similar phase space structures was studied in \cite{berk96,berk97, brei97} for the case of essential dissipation, which is more relevant to our situation. 
Here the} wave resonances do not overlap and the global transport in phase space is suppressed. As the mode grows, most of the particles respond adiabatically to the wave, and only a small  group of resonant particles mix and cause local flattening of the distribution function in phase space within or near the separatrices formed by the waves. However, when  linear dissipation from a background plasma is present, the saturated plateau state becomes unstable and the mode tends to grow explosively. {That} results in the formation and subsequent evolution of long-living{, as compared to the linear growth,} structures in the particle distribution, so called holes (a depletion of particles) and clumps (an excess of particles). These structures represent nonlinear waves of so-called Bernstein--Greene--Kruskal type \cite{BGK} whose frequencies are slightly up- and down-shifted with respect to that of the initial instability. Their subsequent convective motion in phase space is synchronized to the change in wave frequency, thus leading to complex chirping patterns in dynamical spectra of unstable waves, as discussed in more details in \cite{lil10,nyq12,nyq13}. This mechanism is frequently referred as Berk--Breizman model. Formation of holes and clumps is commonly considered as near-threshold phenomena. However recently it has been realized that the same processes may occur far from the instability threshold as well \cite{nyq14}. The underlying physics is that holes and clumps develop from negative energy waves.

Over the years, the hole and clump mechanism has been extensively examined to study and interpret the Alfven wave turbulence driven by high-energy beam ions or alpha particles in toroidal magnetic traps \cite{brei97,lil10,nyq12,nyq13,pinches04,lil09,les10,les13,kosuga12,hole14}; as well as some other MHD processes, such as fishbone  \cite{brei97} and ion-acoustic \cite{les14} instabilities, have been considered. 

In this letter, we report the first laboratory observations of potentially the same mechanism acting in a much higher frequency domain. We study the electron cyclotron (EC) kinetic instabilities of plasma with fast electrons
sustained by high-power microwave radiation under electron cyclotron resonance (ECR) conditions and confined in the mirror magnetic trap. On a decaying phase of a discharge after the microwave heating is switched-off, we find the chirping frequency patterns in the plasma  emission in the EC frequency range that are very similar to those predicted by Berk--Breizman model. The focus of this work was devoted to the study of peculiar time-frequency characteristics of the electromagnetic radiation, which has been made possible only recently with the advent of methods for measuring the electromagnetic field with high temporal resolution. 

It should be noted that research of cyclotron instabilities of nonequilibrium plasma in the laboratory are highly relevant in terms of modeling the physical mechanisms of instabilities in space plasmas \cite{Bingham_2013, VanCompernolle_2015}.  
Using ECR discharge sustained by high-power  microwave radiation (generated with modern gyrotrons) in a magnetic trap allows to increase  the energy deposited into the nonequilibrium plasma component as compared to early experiments. Another  important advantage of modern laboratory study of ECR discharge is an opportunity to recreate very different conditions for excitation and amplification of waves  in the same setup. As a result, generation of pulsed electromagnetic radiation in a laboratory  has much in common with similar processes occurring in the magnetosphere of Earth~\cite{trakh_book_2008,Treumann_2006}, other planets~\cite{Menietti_2012_Jupiter_Saturn}, solar coronal loops \cite{trakh_sun,shal-trah,viktorov_rf_dpr}, stars~\cite{Trigilio_2011_stars}, etc. 

\section{Experimental conditions and results}

The experiments were conducted in the plasma of ECR discharge sustained by gyrotron radiation (frequency 37.5\,GHz, power up to 80\,kW, pulse duration up to 1\,ms)  in the simple axially symmetric open magnetic  trap. 
Microwave radiation traveling through the input teflon window and the matching device is focused in the heating region of the discharge chamber. {The radiation intensity in the focal plane is about 10\,kW/cm$^{2}$ and the average power density is 100\,W/cm$^{3}$.} Discharge chamber is a tube with inner diameter of 38\,mm which is widened in center part by the tube with inner diameter of 72\,mm and 50\,mm length. A discharge chamber is placed in the mirror magnetic trap of the length 225\,mm, produced by pulsed coils with maximum magnetic field strength of 4.3\,T, mirror ratio is about 5, pulse duration $7\un{ms}$. Plasma is created and supported under ECR conditions at the fundamental cyclotron harmonic corresponded to the magnetic  field strength $1.34\un{T}$. The resonance surface is situated between the magnetic mirror  and the center of the discharge chamber. Ambient pressure of a neutral gas (nitrogen or argon) {is about $10^{-6}\un{Torr}$, however it increases up to $10^{-4}-10^{-3}\un{Torr}$ during ECR discharge}. Further details on the setup may be found in \cite{Mansfeld_2007_JETP_maser}.

In the experiments we studied the dynamic spectrum and the intensity of stimulated electromagnetic radiation from the plasma with the use of a broadband horn antenna with a uniform bandwidth in the range from 2 to 20\,GHz and the high-performance oscilloscope Keysight DSA-Z\,594A (analog bandwidth 59\,GHz, sampling rate 160\,GSample/s).
To cover all stages of ECR discharge the length of the recorded waveforms was set to 5\,ms which corresponds to $8\times 10^8$ data points per channel for a single experimental shot.
The dynamic spectra were calculated from the recorded data by short-time Fourier transform windowed with a Hamming window.
Simultaneously,  we measured precipitations of energetic electrons ($>10\un{keV}$)  from the trap ends using a pin-diode detector with time resolution about 1\,ns. As compared to the early studies \cite{avod_whist,shalash_2006,Mansfeld_2007_JETP_maser,shalash_ppcf,viktorov_rf1}, this experiment allowed us to observe the whole picture of instabilities and distinguish details of the radiation spectrum. Similar technique was used in our previous studies \cite{viktorov_rf2,viktorov_rf_dpr,viktorov_EPL}.

\begin{figure*}[t]
	\begin{center}
		\includegraphics[width=165 mm]{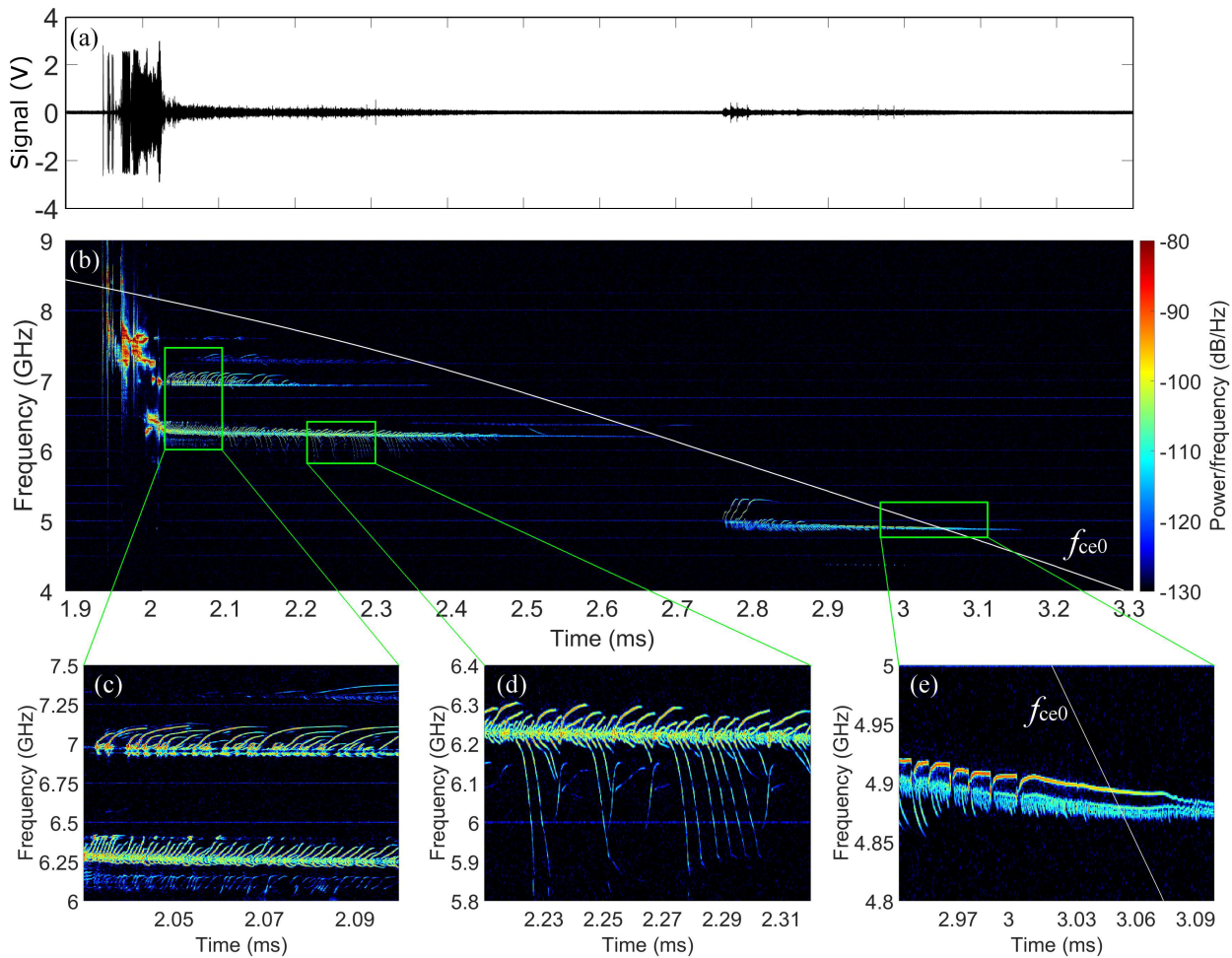}
		\caption{Electric field oscillations in the microwave emission during the nitrogen plasma decay stage (a) and  corresponding dynamic spectrum (b)-(e). Panels (c)-(e) shows the same spectrum in zoom areas. The white line on the spectrograms shows temporal variation of the on-axis electron cyclotron frequency  $f_{\mathrm{ce0}}=\omega_{\mathrm{ce}}(z_{\mathrm{center}})/2\pi$ at the magnetic trap center. The ECR heating is operating during $0-1\un{ms}$ (not shown). Horizontal {dark blue} lines in the specra are rational fractions of Nyquist frequency and related to a digital noise.}
		\label{fig_general}
	\end{center}
\end{figure*}

Operation of an external microwave heating defines three time slots within every experimental shot: the ECR breakdown and the plasma start-up, the developed discharge and the plasma decay stage after the microwave pump switch-off. In the start-up stage, with a duration of about $100\un{\mu s}$, plasma density is small and microwave energy absorbed in a plasma volume is enough to heat electrons up to relativistic energies \cite{viktorov_rf2}. In the developed discharge stage, until the end of the microwave pulse, the plasma density is  by more than two orders of magnitude higher than during the first stage. Here, the two-component electron population is formed containing the cold dense electrons with Maxwellian distribution ($N_\mathrm{c} \sim 10^{13}\un{cm^{-3}}$, $T_\mathrm{c} \sim 300\,\un{eV}$), and less dense component of hot electrons with anisotropic distribution function ($N_\mathrm{h} \sim 10^{11}\un{cm^{-3}}$, mean energy $T_\mathrm{h} \sim 15\,\un{keV}$, although energy tail is stretched up to  $200-300\,\un{keV}$). The third stage starts right after the microwave pulse termination in a decaying plasma, when the temperature and density of the cold fraction decreases rapidly, while the hot electrons are adiabatically confined in the magnetic trap much longer. Therefore, starting from a certain time, the density of the hot component can become comparable to or even higher than the density of the cold component \cite{Mansfeld_2007_JETP_maser}. 

\begin{figure}[t]
	\includegraphics[width=85 mm]{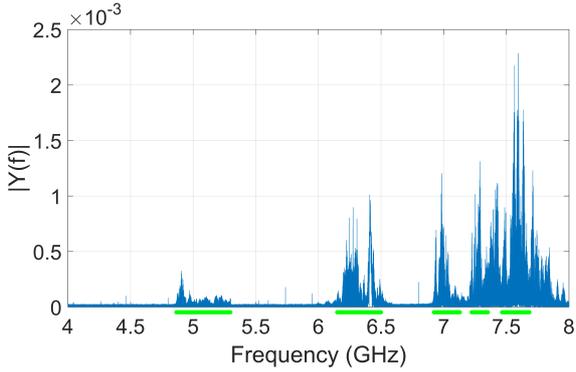}
	\caption{Single-sided amplitude spectrum of the waveform shown in fig.~\ref{fig_general}(a). Frequency bands $4.9-5.3$\,GHz, $6.2-6.5$\,GHz, $6.9-7.1$\,GHz, $7.2-7.3$\,GHz and $7.5-7.6$\,GHz are highlighted by green lines.}
	\label{fft_a020}
\end{figure}

Different stages of pulsed ECR discharge offer the opportunity to simultaneously study wave-particle interactions for essentially different plasma parameters \cite{viktorov_EPL}. However, highly transient spectra with many repeated frequency sweeps have been observed only in a decaying plasma ($N_\mathrm{h} \gtrsim N_\mathrm{c}$) and only with a pronounced delay after the ECR heating switch-off. 
Typical example of such event in a nitrogen discharge is shown in fig.~\ref{fig_general}.  Excitation of such chirping wave packets is observed after the microwave heating switch-off with a delay from 0.1 to 1\,ms and only when ambient magnetic field is decreasing in time. Frequency of the observed microwave emission is always below the electron cyclotron frequency $f_{\mathrm{ce0}}$ at the trap center. The instability development starts with the pulse of continuous broadband emission in the frequency range from 7 to 8 GHz. After that initial pulse which lasts approximately 50\,$\mu$s, the spectrum of emission is significantly changed. The microwave emission is observed only in a few  narrow frequency bands $7.5-7.6$\,GHz, $7.2-7.3$\,GHz, $6.9-7.1$\,GHz, $6.2-6.5$\,GHz and $4.9-5.3$\,GHz, that are independent of all experimental conditions. These frequency bands are clearly seen in a single-sided amplitude spectrum of the whole waveform shown in fig.~\ref{fig_general}(a), see fig.~\ref{fft_a020}.
Within each frequency band the dynamic spectrum represents a set of highly chirped narrowband radiation bursts with both increasing and decreasing frequencies. The bandwidth of this emission inside a single wave packet is about $2\times 10^{-3} f_{\mathrm{ce0}}$.
The duration of the single narrowband spike is about $10\un{\mu s}$; the overall duration of a series of such pulses is up to 1\,ms depending on the experimental conditions. 

An important feature of the band microwave emission  is that in most cases it is observed only when its frequency is below the electron gyrofrequency $f_{\mathrm{ce0}}$ at the trap center. When the magnetic field  decreases such that $f_{\mathrm{ce0}}$ becomes lower than the corresponding frequency band, the microwave emission in  this particular band is not detected anymore. Simultaneously, the microwave emission at lower frequency bands may still be detected or even  triggered on in a new band. 

The fine structure of the emission spectrum in a particular frequency band is changing with the ambient magnetic field decrease. When $f_{\mathrm{ce0}}$ is significantly higher than emission frequency, we observe predominance of rising tones with sweeping frequency that  saturates at more or less constant level for all neighboring pulses. With the decrease of $f_{\mathrm{ce0}}$, falling tones appear in the spectrum. Finally, when $f_{\mathrm{ce0}}$ is close to the frequency of the microwave emission, a narrowband continuous signal without pronounced chirping patterns is observed. Thus, the dynamic bandwidth of the emission is reduced with the decrease of difference between the gyrofrequency and the emission frequency. 

\begin{figure}
	\includegraphics[width=85 mm]{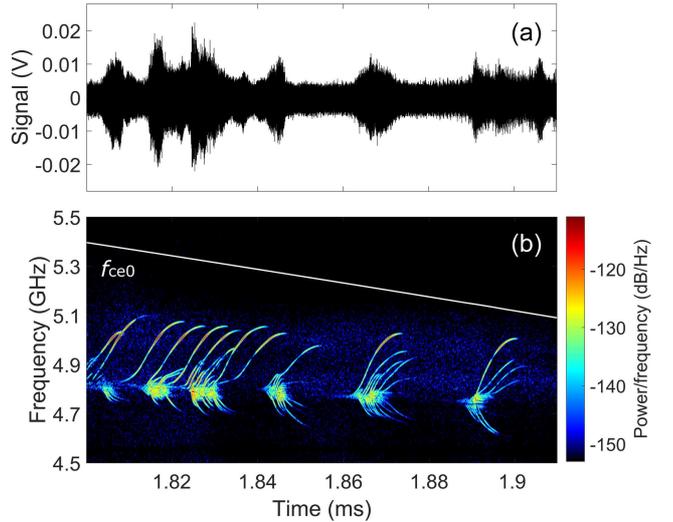}
	\caption{Electric field oscillations (a) and  corresponding dynamic spectrum (b) for the microwave emission during the decay stage of  argon discharge. }
	\label{fig_argon}
\end{figure}

Most of the discussed experiments have been performed in nitrogen plasma. However, an essentially similar dynamics was observed in argon plasma as well. Figure \ref{fig_argon} shows an example of the frequency sweeping in a decaying plasma of argon discharge. Although  less representative statistics is available for argon,    it seems that chirped emissions of the argon discharge are possible in a more narrow parameter range than for the nitrogen plasma. This is related, eventually, to more slow decay of the background plasma with  argon ions as compared to the nitrogen plasma (the latter is dominated with dissociative recombination of molecular ions \cite{Mansfeld_2007_JETP_maser}).

Unlike all other types of instabilities observed  earlier, in the studied case we do not detect the precipitation of hot electrons from the trap that usually accompanies the electromagnetic emission.

\section{Discussion}

The reported dynamic spectra of the plasma emission look very similar to those predicted by Berk--Breizman model in lower frequency domains \cite{lil10,les13,nyq14}. Before comparison to this model, let us summarize shortly our previous knowledge on the kinetic instabilities in a decaying plasma at our  setup.

Such instabilities were first observed as quasi-periodic series of pulsed energetic electron precipitations that appeared with some delay after the switching off the heating power \cite{Mansfeld_2007_JETP_maser}, later rough estimations of plasma EC emission spectrum were obtained using microwave detectors \cite{viktorov_rf1}. The instability  was interpreted as a result of resonant interaction at the fundamental cyclotron harmonic between the energetic electrons and the slow extraordinary waves propagating in a rarefied plasma across the external magnetic field at frequencies below the local gyrofrequency \cite{Mansfeld_2007_JETP_maser}. The linear growth rate  was estimated as $\gamma_{\mathrm{L}}\sim 5\times 10^7$~s$^{-1}$.
The delay between the start of the decay and triggering of the instability was explained by a polarization depression effect of the background (more dense and cold) plasma.  In an unbounded dense plasma, the slow extraordinary wave is nearly circular polarized with the rotation direction opposite to the electron gyration direction, therefore, the wave-electron interaction is strongly weakened. However, in a rarefied plasma satisfying  the condition 
\begin{equation}\label{eq:rare}
\omega_{\mathrm{pe}}^2\ll\omega_{\mathrm{ce}}^2-\omega^2\sim \omega_{\mathrm{ce}}^2\; \varepsilon/mc^2
\end{equation}
the extraordinary wave is polarized almost linearly, which ensures an effective interaction with resonant electrons. Here $\omega$, $\omega_{\mathrm{pe}}$, $\omega_{\mathrm{ce}}$ are, correspondingly, the wave, electron Langmuir and cyclotron frequencies calculated at the trap center, and $\varepsilon$ is the kinetic energy of a radiating electron. Hence, the cyclotron instability growth rate increases with the decreasing background plasma density.
With $ck_{||}/\omega_{\mathrm{ce}}\ll\sqrt{\varepsilon/mc^2}$, the interaction occurs at the relativistically downshifted electron gyrofrequency 
\begin{equation}\label{eq:res}
\omega\approx\omega_{\mathrm{ce}}/(\varepsilon/mc^2+1)\,.
\end{equation} 

The sequences of pulsed bursts at the nonlinear instability phase was explained in terms of a cyclotron maser model with a fast decrease of electromagnetic energy losses  \cite{shalash_2006}. It has been shown that, even in the absence of a continuously acting source of nonequilibrium particles
(inversion of the medium), the instability condition, $\gamma_{\mathrm{L}}>\gamma_{\mathrm{d}}$ with $\gamma_{\mathrm{d}}\approx\nu_{\mathrm{ei}}$ being the wave damping rate due to collisional absorption by the background plasma, is recovered after each individual burst due to monotonic decrease of $\nu_{\mathrm{ei}}(t)$ during the plasma decay.
Therefore, the dynamics of the background  plasma is very essential for the proposed model since decreasing wave losses in the background plasma actually pump the maser instability. Unfortunately,  experimental measurements of the background plasma parameters have been possible only for the steady state ECR discharge and no reliable data are available describing the decay phase. So we reconstruct the electron density and temperature during the plasma decay basing on the particle and energy balance equations solved with initial conditions corresponded to the measured parameters of the ECR discharge \cite{Mansfeld_2007_JETP_maser,shalash_ppcf}.

\begin{figure}
\includegraphics[width=80 mm]{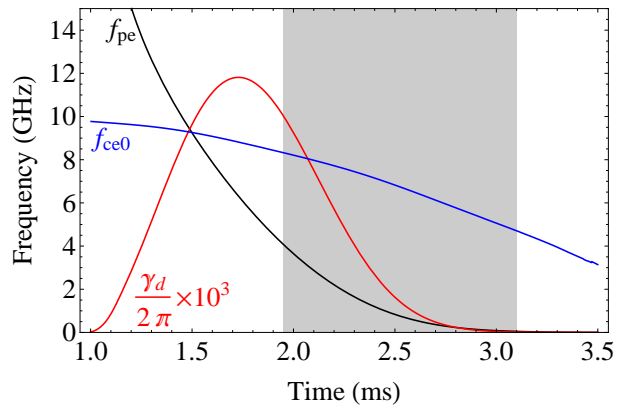}
\caption{Evolution of the plasma frequencies  $f_{\mathrm{pe}}$,  $f_{\mathrm{ce}}$, the background collision rate $\gamma_{\mathrm{d}}$ during the plasma decay. The time interval where the frequency sweeps are observed is shown in gray.
Used model is discussed in \cite{Mansfeld_2007_JETP_maser, shalash_ppcf}; modeling is performed for the following  initial conditions relevant to the experimental shot shown in fig.~\ref{fig_general}:
$N_\mathrm{e}(0)=N_\mathrm{i}(0)=2\times 10^{13}\;\mathrm{cm}^{-3}$, $N_\mathrm{m}(0)=N_\mathrm{m}^{+}(0)=0$, $T_\mathrm{e}(0)=300\;\mathrm{eV}$. A constant flux of puffed gas is assumed, $F=1.5\times10^{16}\;\mathrm{cm}^{-3}\;\mathrm{s}^{-1}$. Energy of the fast electrons is $\varepsilon\approx 165$\,keV.  }
\label{fig_decay}
\end{figure}

In the present work we derive dynamic spectrum and observe its fine structure, which was not possible before.
In fig.~\ref{fig_decay} we show the typical evolution of  plasma frequencies and collisional rates during the plasma decay, corresponded to case shown in fig.~\ref{fig_general}. One can see that instability starts in $t_*\approx2\;\mathrm{ms}$ when condition (\ref{eq:rare}) becomes full-filled, $\omega_{\mathrm{pe}}/\omega_{\mathrm{ce}}\sim 0.5$. At this moment, the  electron temperature $T_e$ is about $0.1\;\mathrm{eV}$,  the  background plasma
density is $N_e\sim 2\times 10^{11}\;\mathrm{cm}^{-3}$, and the characteristic growth rate is
$\gamma_{\mathrm{L}}\approx\gamma_{\mathrm{d}}( t_*)\approx 6.3\times10^7\;\textrm{s}^{-1}$.
Dynamic spectrum of the plasma emission can be used to estimate energies of emitting electrons from the resonant condition (\ref{eq:res}). For the case shown in fig.~\ref{fig_general}, maximum detuning $\omega_{\mathrm{ce}}/\omega\approx 1.32$ is observed at the initial phase of the instability development just after the first broadband pulse. This indicates that electrons are accelerated up to 165\,keV. In a consequent series of repetitive chirped pulses the radiation frequency approaches the gyrofrequency, $\omega_{\mathrm{ce}}/\omega\to 1$, so the energy of resonant electrons is decreasing. 

Isolated narrow frequency bands observed in the microwave emission spectrum may be explained by excitation of individual modes of a bounded plasma in a discharge chamber. As a rough estimate supporting this idea, one may consider first several vacuum modes of the cylindrical plasma 
chamber. Such  waveguide has cut-off frequencies 5.08\,GHz, 6.54\,GHz, 6.80\,GHz, 7.07\,GHz, 7.32\,GHz etc, that are very close to the observed frequency bands (fig.~\ref{fft_a020}). Of course, a more detailed analysis of excitation of a plasma-filled waveguide is needed for quantitative description of a linear phase of the instability. A similar problem was studied in \cite{Vorgul_plasma_modes}.

Our previous attempt to explain the observed complex dynamics demonstrated by the cyclotron maser in a decaying plasma discharge was based on the idea of self-modulation of a maser due to interference of two counter-propagating degenerate unstable waves resulting in spatial modulation of amplification \cite{shalash_ppcf}. However this model can not describe frequency sweeps resolved in the reported experiments.

The ``holes and clumps'' paradigm, as formulated by Berk and  Breizman, seems to be the most suitable model to explain our data. Following \cite{berk97} one can isolate one spectral component and model it by a Bernstein--Greene--Kruskal wave  to obtain the time evolution of one chirping event. Recall that $\gamma_{\mathrm{L}}$ is the linear growth rate in the absence of fast particle collisions and external dissipation, and $\gamma_{\mathrm{d}}$ is the  damping rate defined by collisional absorption of waves by the background plasma.  Then, in the collisionless limit for resonant fast particles,  a bounce average of  Maxwell--Vlasov kinetic equation yields the saturation level $\omega_{\mathrm{b}}$ and the frequency shift $\delta\omega$ of an unstable mode  as
\begin{equation}
\label{eq:bb}
\delta\omega\approx \frac{16\sqrt2}{3\sqrt{3}\pi^2}\,\gamma_{\mathrm{L}}\sqrt{\gamma_{\mathrm{d}} t}\,,\quad\omega_{{b}}\approx \frac{16}{3\pi^2} \,\gamma_{\mathrm{L}}\,,
\end{equation}
where  the bounce frequency of particles that are deeply trapped in the wave  potential, $\omega_{\mathrm{b}}$, is used as a measure of the electric field amplitude.
{These} expressions are valid for 
\begin{equation}
|\upd\omega_{\mathrm{b}}/\upd t|\ll|\upd{\delta\omega}/\upd t|\ll\omega_{\mathrm{b}}^2\sim\gamma_{\mathrm{L}}^2,
\end{equation}
i.e. in a regime where perturbation of the passing particle distribution is negligible. Although these analytical results were originally obtained for purely electrostatic and one-dimensional bump-on-tail instability, it gives reasonable qualitative estimates for more complex electromagnetic problems, see e.g. \cite{les10}.  {Similar results were obtained before for the particular case of cyclotron instability of whistler waves in Earth magnetosphere: $\omega_{\mathrm{b}}\approx(32/3\pi)\gamma_{\mathrm{L}} $ in \cite{Trakh_book_1984} and $\omega_{\mathrm{b}}^2\approx2\pi\:\dervi{\delta\omega}{t} $ in \cite{trakh4}.}

\begin{figure}
\includegraphics[width=85 mm]{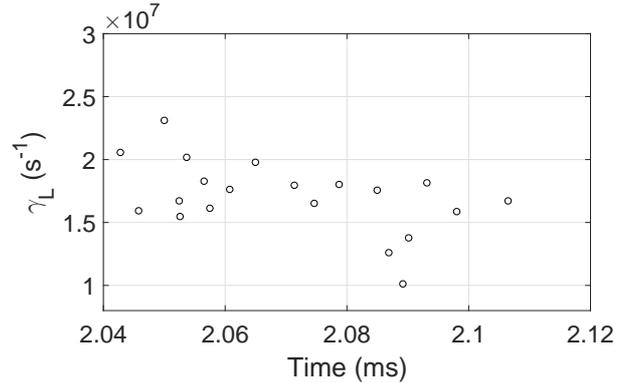}
\caption{The lower boundary for the linear growth rate $\gamma_{\mathrm{L}}$ compatible with Berk--Briezman model for the high frequency part of the dynamic spectrum shown in fig.~\ref{fig_general}(c).}
\label{fig_fit_A}
\end{figure}

Following this logic and understanding that more rigorous analysis should follow, we apply eqs.~(\ref{eq:bb}) to our data. Fitting the measured frequency sweeps $\delta\omega=\sqrt{A\:t}$ by constant $A$, one may estimate the upper boundary for the dissipation $\gamma_{\mathrm{d}}$, or, equivalently, the lower boundary for the linear growth rate $\gamma_{\mathrm{L}}$ compatible with Berk--Briezman model. Indeed, the first equation (\ref{eq:bb}) implies that $\gamma_{\mathrm{L}}^2\gamma_{\mathrm{d}}\approx5A$, and instability condition requires $\gamma_{\mathrm{L}}\geqslant\gamma_{\mathrm{d}}$. Evidently, the above mentioned boundaries correspond to the instability threshold,
$\gamma_{\mathrm{d}}= \gamma_{\mathrm{L}}\approx\sqrt[3]{5A}$.
In the experiments we obtain typical values  $A=(0.4-2)\times 10^{21}$\,s$^{-3}$, what corresponds 
$\gamma_{\mathrm{L}}\approx(1-2.5)\times 10^7$\,s$^{-1}$. 
For example, fig.~\ref{fig_fit_A} shows the evolution of the growth rate obtained by this technique to the spectrogram (c) in fig.~\ref{fig_general}. Here each point corresponds to a separate chirping burst. Note  that this simple estimate, that neglects all geometrical factors related to our particular modes, results in slightly underestimated instability growth rates, but still is in a relatively good agreement with our previous calculations.

{
For the slow extraordinary waves propagating transverse to the external magnetic field in a rarefied plasma the bounce frequency may be approximated as \cite{Trakh_book_1984}
\begin{equation} 
\omega_{\mathrm{b}}^2\approx {\frac{v_\perp}c\frac {E_\sim}{B_0}} \frac{\omega_{\mathrm{ce}}^2} {\gamma^2}\,,
\end{equation}
here $E_\sim$ is a wave electric field, $B_0$ is the external magnetic field strength, and $v_\perp$ is the transverse electron velocity. 
Then from the second expression in~(\ref{eq:bb}) one can estimate the characteristic level of saturated electric field as $E_\sim\approx 2-3$\,V/cm for 60\,keV electrons. 

}	

The above preliminary conclusions need further verification by full non-linear solution of a kinetic equation for the resonant particles, the work to be published separately. Note that the asymmetry of the ascending and  descending frequency sweeps suggests an important role of the drag force (either collisional, or turbulent) acting on the resonant particles \cite{lil10,les13,nyq13,lil09}. Therefore, a realistic collision operator with the ``drag'' and ``diffusion'' parts should be taken into account in the kinetic modeling.

\section{Summary}

In this paper the kinetic EC  instabilities of two-component plasma  sustained by high-power microwave radiation under ECR conditions and confined in the mirror magnetic trap were studied. The chirping frequency patterns in the plasma EC emission, which are very similar to those predicted by Berk--Breizman model, were observed during plasma decay stage, characterized by high relative density of the fast electrons. The fine structure of dynamic spectra of EC emission was investigated with high temporal resolution. The analysis of the experimental data allowed to estimate energies of emitting electrons (165\,keV) and the range of instability growth rates (up to $6\times 10^7$\,s$^{-1}$). 

This paper may be of interest in the context of a laboratory modeling of non-stationary processes of wave-particle interactions in space plasma, since there are a lot of open questions about the origin of some types of emissions in space cyclotron masers, especially mechanisms of fine spectral structure \cite{Treumann_2006}.  Although the concept of electron holes in a phase space was occasionally used, e.g. for interpretation of emissions in the Earth's auroral upward and downward current regions \cite{akr_holes1,akr_holes2}, Jovian S-burst emission \cite{astr15}, the self-consistent theory of fast frequency events similar to those of toroidal Alfven eigenmodes in tokamaks is still absent. We hope that the reported dynamic spectra with frequency sweeping may eventually push the application of ``holes and clumps'' paradigm into a new realm.

\acknowledgments
The work has been supported by RFBR (grants No. 15--32--20770, 16--32--60056). 
The authors are grateful to Professor Andrei Demekhov for fruitful discussions and Keysight Technologies Inc. for its technical support.

\end{document}